\documentclass[twocolumn,showpacs,APS]{revtex4}  
\usepackage{amssymb}
\usepackage{amsmath}
\usepackage{graphicx}
\usepackage{bm}

\renewcommand{\d}{\text{d}}

\begin{document}

\title{Nonequilibrium fluctuations of an interface under shear}
\author{Marine Thi\'ebaud    and Thomas Bickel}    
\email{thomas.bickel@u-bordeaux1.fr}         
\affiliation{ CPMOH, Universit\'{e} de Bordeaux \& CNRS (UMR 5798) \\
351 cours de la Lib\'eration, 33405 Talence, France}
\date{September 4, 2009}

\begin{abstract}

The steady state properties of an interface in a stationary Couette flow are addressed within the framework of fluctuating hydrodynamics.
Our study reveals that thermal fluctuations are driven out of equilibrium by an effective shear rate that differs from the applied one.
In agreement with experiments, we find that the mean square displacement of the interface is strongly reduced by the flow. 
We also show that  nonequilibrium fluctuations present a certain degree of universality
in the sense that all features of the fluids can be factorized into a single control parameter. 
Finally, the results are discussed in the light of recent experimental and numerical studies.

\end{abstract}


\maketitle

\section{Introduction}
\label{intro}

Soft matter systems driven far from equilibrium by a shear flow manifest striking properties as a result of their sensitivity to external fields~\cite{catesbook}.
For instance, hydrodynamics may either enhance or suppress coarsening processes~\cite{onukiJPCM97}. 
Coupling  with the flow can also lead to the emergence of new, shear-induced phases that are exclusively out-of-equilibrium 
structures~\cite{diatJPII93,tanakaPRL96}, whereas spatiotemporal oscillations and rheochaos are observed in shear-banding systems~\cite{becuPRL04,fieldingSoft07}.

Despite substantial progress, a fundamental understanding of complex fluids under shear remains a challenging question. 
The first reason lies in the nature of the coupling between structure and flow. Since the local structure of soft materials is readily reorganized by an external flow, it has in turn a significant impact on the flow itself. The complexity of this feedback mechanism makes that most theoretical studies are based -- to a variable extent -- on phenomenological models~\cite{catesbook,onukiJPCM97}.
The second point is that our understanding of nonequilibrium steady states (NESS) is yet far from complete.  Recently, there have been several attempts to construct a unified framework to describe equilibrium and nonequilibrium phenomena~\cite{sevickAnRev08}.
In particular, fluctuations theorems~\cite{baulePRL08} or extended fluctuation-dissipation relations~\cite{speckPRE09} have been suggested for complex fluids under shear.
But  these relations still remain at the conceptual level and the link with experimental observables -- power spectra, correlation functions -- has not been clarified yet.

In this paper, we investigate the statistical properties of a liquid-liquid interface in a stationary flow. 
We argue that this system is sophisticated enough to capture the relevant features of NESS but remains simple enough to be addressed analytically. 
Applications of this issue might be expected for instance in microfluidics, since interfacial phenomena are inevitably enhanced as the size of the system is 
reduced. Still, the question of interface fluctuations under shear has received only little attention so far.
From a fundamental viewpoint, equilibrium properties of liquid-liquid interfaces are now well established~\cite{langevinbook}, and
experimental as well as numerical studies have validated the capillary wave model down to almost molecular scales~\cite{fradinNat00,delgadoPRL08}. 
This topic has experienced a renewed interest in recent years with the discovery that phase-separated colloid-polymer mixtures may have an extremely low interfacial tension~\cite{aartsSci04}. Interface fluctuations can then be analyzed in real time and space using elementary video-microscopy techniques. The versatility of this method has  allowed Derks and collaborators to study the statistical properties of an interface exposed to a shear flow~\cite{derksPRL06}. They found that the coupling with the flow  leads to
a strong reduction of thermal fluctuations, while the correlation length increases. But this second point is in disagreement with recent Monte Carlo simulations of a driven Ising model~\cite{smithPRL08}, therefore raising the fundamental question of what features of interfaces under shear are actually universal.

Statistical fluctuations of an interface are driven by the  random forces that spontaneously occur in the bulk~\cite{grantPRA83}. 
To describe nonequilibrium properties, the main issue is thus to properly account for the coupling between the bulk and the interface. 
This can be achieved on the basis of fluctuating hydrodynamics (FH)~\cite{landaups2}. 
Indeed, FH has been successfully applied   to various NESS situations, and experimental validations have been obtained, e.g., for  fluids in a temperature gradient~\cite{ortizbook}.
The purpose of this paper is to apply FH to an interface under shear.

We shall proceed as follow. In Sec.~\ref{fluct}, we explain how an equation of motion for the interface can be obtained from
FH. We show that the distortion of capillary waves by the flow leads to a mode-coupling equation
that is discussed in Sec.~\ref{mode}. This allows us to extract  NESS properties  under a stationary flow in Sec.~\ref{neq}.  
In particular, we find  that the fluctuations are smoothed out by the flow.
The results are then discussed in Sec.\ref{disc}  in the light of recent experimental and numerical data  available in the literature. Finally, we conclude the paper with a short summary of our results. For the sake of clarity, details of the algebra are presented in Apps.~\ref{app1}--\ref{app4}.

\section{Hydrodynamic formulation}
\label{fluct}

We first set up  a hydrodynamic theory to account for the  coupling between the surface and the bulk. 
Following the usual hypothesis of capillary wave theory,
we assume the existence of an intrinsic interface separating two immiscible fluids. 
For moderate deformations around the $xOy$ horizontal plane, the position of the interface can be described by a single valued function $z=h(x,y,t)$. 
Along this article, properties of the upper (resp.~lower) fluid are labeled with
the subscript~$i=1$ (resp.~$i=2$).
Each phase is characterized by its mass density $\rho_i$ and its viscosity $\eta_i$.
We also define $ \bar{\eta}=(\eta_1 + \eta_2)/2$ the mean viscosity and $\Delta \rho =  \rho_2 - \rho_1 >0$ the mass density difference.
The surface is further characterized by the interfacial tension $\sigma$ and the capillary length
$l_c=\sqrt{ \sigma/(\Delta \rho  g )}$, with  $g$ the gravitational acceleration.

\begin{figure}
\includegraphics[width=80mm]{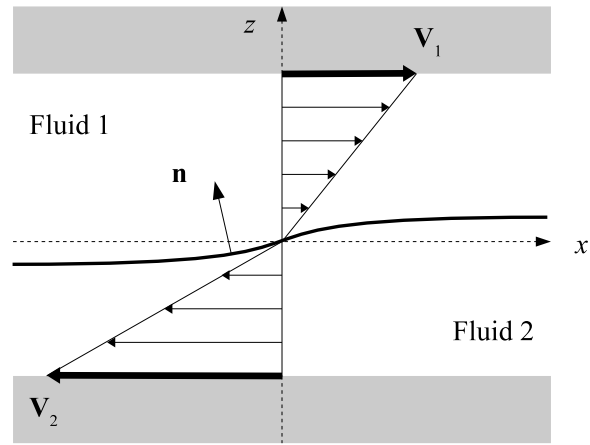}
\caption{Schematic representation of the system. The unit vector $\mathbf{n}$ is normal to the interface and is pointing towards the upper fluid.}
\label{schema}
\end{figure}

The system is schematically drawn in Fig.~\ref{schema}. The average position of the interface is $z=0$.
It is confined between to walls, the thickness of each fluid layer being $L_1$ and $L_2$ with $L=L_1+L_2$.
A planar Couette flow  is induced  by moving the walls of the shear cell at constant velocity along the $x$ direction.
We define $x$, $y$ and $z$ respectively as the velocity, vorticity and velocity gradient directions.
Assuming that the no-slip condition applies on the walls of the cell, the fluid velocity $\mathbf{v}$ satisfies $\mathbf{v}(x,y,L_1)=V_1\mathbf{e}_x$ and $\mathbf{v}(x,y,-L_2)=-V_2\mathbf{e}_x$. 
The total shear rate is then
 \begin{equation}
\dot{\gamma}=\frac{\dot{\gamma}_1  L_1+\dot{\gamma}_2  L_2}{L_1+L_2}     \ ,
 \label{gammatot}
 \end{equation}
where we define $\dot{\gamma}_i = V_i / L_i$ the shear rate in each phase.
Without loss of generality, we  assume  in the following that $V_1$, $L_1$, $V_2$ and $L_2$ are chosen so that the plane of zero shear coincides with the average position of the interface. Other situations can be deduced thanks to a Galilean transformation.

For usual fluids at room temperature, thermal fluctuations occur in the overdamped regime of capillary waves.
In order to describe nonequilibrium effects, the analysis is performed within the framework of fluctuating hydrodynamics~\cite{landaups2,ortizbook}.
The starting point is the stochastic version of the Stokes equation 
 \begin{equation}
 \eta_i \nabla^2 \mathbf{v} - \bm{\nabla} p + \rho_i \mathbf{g} +  \bm{\nabla} \cdot \mathsf{s}  = \mathbf{0}  \ ,
 \label{stokes}
 \end{equation}
with $\mathbf{v}$ the velocity field, $p$ the pressure, and $\mathbf{g} =-g \mathbf{e}_z $. Eq.~(\ref{stokes}) is solved
together with the incompressibility condition 
\begin{equation}
\bm{\nabla} \cdot  \mathbf{v} =0 \ .
 \label{incomp}
 \end{equation}
 Thermal fluctuations are accounted for through the random part of the stress tensor $\mathsf{s}$. Its components $s_{\mu \nu}$ (with $\mu, \nu=x$, $y$, or $z$) are stochastic forces that stem from the microscopic degrees of freedom of the fluids. Close to equilibrium their correlations  are given by the fluctuation-dissipation theorem, but such a relation is not expected to hold beyond the regime of linear response. Here however, we shall take advantage of the separation of time scales between the collective modes under study -- the fluctuations of the interface -- and the molecular scales of the heat bath --  the fluid constituants. 
The relaxation of an interface is characterized by the capillary time $\tau_c=2 \bar{\eta} l_c / \sigma$; it ranges from milliseconds for usual  interfaces ($\sigma \approx 10^{-2}$~mN/m)  to a few seconds for ultra-soft  interfaces ($\sigma \approx 10^{-6}$~mN/m). On the other hand, a particle of fluid on either side of the interface diffuses over its own diameter $a$ on a time-scale $\tau_b = \eta_i a^3 / (k_BT)$. This time scale  is at most of the order of $\tau_b\approx 10^{-3}$~s for particulate fluids with $a \approx 100$~nm~\cite{derksPRL06}, but it is several orders of magnitude smaller for molecular fluids.
Since the relevant regime considered in this work corresponds to $\dot{\gamma} \tau_c \sim 1$, the applied shear rate is  too small to significantly affect the thermal motion of individual particles. In this small Peclet number limit ($\dot{\gamma} \tau_b \ll 1$), the statistics of the heat bath is not affected by the flow  and  we can  assume that the  stochastic variables have zero mean value and correlations given by~\cite{landaups2,ortizbook}
 \begin{eqnarray}
\left\langle s_{\mu \nu} \left( \mathbf{r},t \right)   s_{\mu' \nu'} \left( \mathbf{r}',t' \right)  \right\rangle  
 =  2  k_B T \eta_i   \left( \delta_{\mu \mu'}\delta_{\nu \nu'} + \delta_{\mu \nu'}\delta_{\nu \mu'} \right) \nonumber \\
\times  \delta \left( \mathbf{r} - \mathbf{r}'  \right) \delta (t-t')  \ ,
 \label{randomten}
 \end{eqnarray}
with $k_B$ the Boltzmann's constant and $T$ the  temperature.

Eqs.~(\ref{stokes}) and~(\ref{incomp}) are then solved above and below the interface, and the solutions  matched with the appropriate boundary conditions. The latter have to be enforced at the instantaneous location of the interface $z=h(x,y,t)$.
Explicitly, we need to express the continuity of the velocity
 \begin{equation}
\big[ \mathbf{v} \big]_h = 0 \ ,
 \label{velocity}
 \end{equation}
as well as the continuity of the stress
 \begin{equation}
\big[ \mathsf{T} \big]_h \cdot \mathbf{n} = \sigma \,  \mathbf{n} \left( \bm{\nabla} \cdot \mathbf{n} \right) \ ,
 \label{stress}
 \end{equation}
with the notation $\left[ f \right]_{z_0} = f(z_0^+)-f(z_0^-)$, the limit being taken respectively from above and from below.
In Eq.~(\ref{stress}), $\mathsf{T}=\mathsf{t}+\mathsf{s}$ is the \emph{total} stress tensor~\cite{haugeJSP73,grantPRA83}. The components of  $\mathsf{t}$ read $t_{ \mu \nu}=-p\delta_{ \mu ,\nu} + \eta_i \left( \partial_\mu v_{\nu} + \partial _\nu v_{\mu}    \right)$, with $\mu,\nu=x,y, \ \text{or} \ z$; the random part $\mathsf{s}$ is defined in Eq.~(\ref{stokes}).
The unit vector $ \mathbf{n}$ is normal to the surface, pointing towards the upper fluid. It depends on the local conformation of the interface~\cite{kamienRMP02}
 \begin{equation}
 \mathbf{n}  = \frac{1}{\sqrt{1+ (\bm{\nabla} h)^2}} 
 \begin{pmatrix}
 -\partial_x h \\  -\partial_y h \\ 1
 \end{pmatrix}  \ .
 \label{normal}
 \end{equation}

Finally, once the velocity field is fully characterized, an equation of motion is  obtained thanks to the kinematic relation
 \begin{equation}
 \partial_t h   + \mathbf{v}_{\parallel} \cdot \bm{\nabla}_{\parallel} h = v_z  \ ,
 \label{kinematic}
 \end{equation}
the velocity being evaluated at $z=h$.
In this equation, the parallel components  of a  vector field $\mathbf{f}=(f_x,f_y,f_z)$ are defined as $\mathbf{f}_{\parallel}=(f_x,f_y)$.
For the sake of completeness, the derivation of the boundary conditions~(\ref{stress}) and~(\ref{kinematic}) is reminded in Appendix~\ref{app1}.

\section{The interface equation}
\label{mode}

It has been appreciated for a long time that interfacial fluctuations can be distorted by an external flow~\cite{onukiAP79,bruinsmaPRA92}. 
The description of this effect requires to go beyond the linear analysis generally used to derive the dispersion relation for capillary waves. 
To proceed, we follow a recursive scheme that has proved its worth, e.g., in the context of 
polymer-membrane interactions~\cite{bickelEPJE02}. 
We assume that the deformation can be written $h(x,y,t)=\varepsilon u(x,y,t)$, with $u(x,y,t)\sim \mathcal{O} (1)$.
The dimensionless parameter that governs this small-gradient expansion is $\varepsilon = \sqrt{k_BT / (\sigma l_c^2)}$.
The velocity and pressure fields are then expressed as
 \begin{eqnarray*}
\mathbf{v} &=& \mathbf{v}^{(0)}  + \varepsilon \mathbf{v}^{(1)} + \varepsilon^2 \mathbf{v}^{(2)} + \ldots \\
p &=& p^{(0)}  + \varepsilon p^{(1)} + \varepsilon^2 p^{(2)} + \ldots
\end{eqnarray*}
The lowest-order term is simply the Couette flow solution for a planar interface: $\mathbf{v}^{(0)}(\mathbf{r})=\dot{\gamma}_1 z \mathbf{e}_x$ if $z \geq 0$, and $\mathbf{v}^{(0)}(\mathbf{r})=\dot{\gamma}_2 z \mathbf{e}_x$ if $z \leq 0$. 

For $n \geq 1$, the linearity of the Stokes equation   implies that each term of the series obeys  Eq.~(\ref{stokes}).
The coupling between successive orders originates from the boundary conditions. Indeed, the latter have to be enforced at the position of the interface.
It is therefore expected  that the Taylor expansion of the $n^{th}$-order field, when evaluated at $z=\varepsilon u(x,y,t)$, involves contributions of all orders 
$k \geq n$.
To be more explicit, consider for instance the velocity field. Up to second order, it is given by
 \begin{eqnarray}
\mathbf{v} (\varepsilon u) 
&=&\mathbf{v}^{(0)} (0)  +\varepsilon \left[\mathbf{v}^{(1)} (0) + u \partial_z  \mathbf{v}^{(0)} (0) \right]  \nonumber \\
& +&\varepsilon^2 \left[ \mathbf{v}^{(2)} (0) + u \partial_z  \mathbf{v}^{(1)} (0) +  \frac{u^2}{2}\partial_z^2  \mathbf{v}^{(0)} (0) \right] + \ldots  \nonumber
 \end{eqnarray}
[Note that we have only  written the $z$-dependence of $\mathbf{v}(x,y,z,t)$].
We proceed likewise for all fields in Eq.~(\ref{velocity}) -- (\ref{kinematic}).
Clearly, the Taylor expansion of the boundary conditions involves several second-order contributions, making the algebra quite cumbersome. 
Moreover, the study of the fluctuations requires us to solve the full three-dimensional problem.
We thus defer the technical details to App.~\ref{app2} and~\ref{app3}, and we now focus the discussion on the main outcomes.

Since the  problem is invariant by translation parallel to  the horizontal plane, it is  natural to switch  to the (2D) Fourier representation
$h (\mathbf{q},t)=\int \d^2 \mathbf{r}{\rho}\exp [-i\mathbf{q} \cdot \mathbf{r}]h(\mathbf{r},t)$,
with $\mathbf{r}=(x,y)$ and $\mathbf{q}=(q_{x},q_{y})$.
We also define the norm  of the wave vector $q = \vert \mathbf{q} \vert = (q_x^2+q_y^2)^{1/2}$.
After some algebra, we find that the relaxation of a fluctuation mode with wave vector $\mathbf{q}$ follows a \emph{mode-coupling equation} 
 \begin{align}
 \partial_t h=- \frac{1}{\tau_q} h(\mathbf{q},t)   -  i\dot{\gamma}_{\mathit{eff}} \int \frac{\d^2 \mathbf{k}}{(2\pi)^2} &
k_x h(\mathbf{k}  ,t) h(\mathbf{q-k},t)  \nonumber \\ 
&+ \varphi (\mathbf{q},t)   \ .
 \label{mceq}
 \end{align}
This equation constitutes the first main result of this paper. It involves a number of contributions that we now discuss.
First, $\tau_q=4 \bar{\eta}q/(\sigma( q^2 + l_c^{-2}))$ is the equilibrium relaxation time of the interface. This result shows that  there is no direct coupling at linear order (at least in the viscous regime). 
Advection of the interface by the flow takes the form of a convolution between all modes with an effective shear rate
 \begin{equation}
\dot{\gamma}_{\mathit{eff}} = \frac{ \eta_1 \dot{\gamma}_1 +\eta_2 \dot{\gamma}_2}{ \eta_1 +  \eta_2} \ .
 \label{gammaeff}
 \end{equation}
Note that the second-order term does not depend on the elastic properties of the interface.

The special feature of Eqs.~(\ref{mceq}) and (\ref{gammaeff}) is that the effective shear rate $\dot{\gamma}_{\mathit{eff}}$  felt by the interface differs from the applied shear rate  $\dot{\gamma}$ defined in Eq.~(\ref{gammatot}). Although the latter is set by the geometry of the system,
the former  is a dynamical quantity in the sense that it depends on the viscosities of both fluids.
However, $\dot{\gamma}_{\mathit{eff}} $ and $\dot{\gamma}$ cannot be tuned independently since the continuity condition~(\ref{stress}) for  tangential forces
requires $\eta_1 \dot{\gamma}_1 = \eta_2 \dot{\gamma}_2$ (see App.~\ref{app2}), and then
 \begin{equation}
 \left(   \frac{L_1}{\eta_1}  + \frac{L_2}{\eta_2}\right) \dot{\gamma}_{\mathit{eff}} =  \frac{L}{\bar{\eta}} \, \dot{\gamma} \ .
 \label{relgamma}
 \end{equation}
This relation implies that the effective shear rate
can in principle be tuned in the range $0 \leq \dot{\gamma}_{\mathit{eff}} \leq 2\, \dot{\gamma}$ by adjusting the experimental conditions.
It is only when $\eta_1= \eta_2$ that both shear rates coincide.

Thermal fluctuations are accounted for through the white noise $\varphi (\mathbf{q},t)$.
This contribution stems from the random part of the stress tensor (see Appendix~\ref{app3}). 
We find  that $\varphi (\mathbf{q},t)$ simply follows the equilibrium distribution
with zero mean value and 
 \begin{equation}
\langle \varphi (\mathbf{q},t)\varphi (\mathbf{q}',t')\rangle = 
 \frac{k_BT}{2\bar{\eta} q}   \delta (t-t')  (2\pi)^2 \delta ( \mathbf{q}+\mathbf{q}')   \ .
 \label{noisecorrel}
 \end{equation}
Even though the system is driven far from equilibrium, there is no coupling between the noise and the external flow
 (at least up to $\mathcal{O} (\varepsilon^2)$)~\cite{rem1}.

At this point, it should be mentioned that corrections similar to Eq.~(\ref{mceq}) have been suggested in the context of sheared smectic phases (see~\cite{marlowPRE02} and references therein), or in the description of coarsening under shear~\cite{brayPRE01a,brayPRE01b}. 
The argument commonly invoked is the following. Since the deformation $h(x,y,t)$ is dragged along the $x$ direction by the external flow $\mathbf{v}^{(0)}=v^{(0)}\mathbf{e}_x$, the time derivative involved in the equation of motion can simply be replaced by the total derivative
 \begin{equation*}
\partial_t h \rightarrow \partial_t h + v^{(0)} \partial_x h \ ,
 \end{equation*}
where $v^{(0)}$ is evaluated at $z=h$. Of course if the slope of the velocity  is continuous, i.e. if $\eta_1=\eta_2$, there is no ambiguity: the total derivative then reads $\partial_t h + \dot{\gamma} h \partial_x h$, yielding to the convolution integral in Fourier space.
However, the situation is more subtle when $\eta_1\neq \eta_2$. With the same argument, the velocity evaluated at $z=h$ would be 
$v^{(0)} = \dot{\gamma}_1h$ if $h\geq 0$, whereas it would take the value $v^{(0)} = \dot{\gamma}_2 h$ if $h\leq 0$. Clearly, there is an inconsistency in the reasoning.
The only way to remove the indeterminacy is to follow the procedure detailed in Appendix~\ref{app2}. We emphasize in particular that the various second-order contributions in Eq.~(\ref{velocity}) -- (\ref{kinematic}) are all equally important to give the perfectly symmetric relation Eq.~(\ref{gammaeff}).

\section{Nonequilibrium fluctuations}
\label{neq}

NESS properties of the interface are then extracted from
the equation of motion~(\ref{mceq}) together with the noise correlations~(\ref{noisecorrel}).
Unfortunately, this class of nonlinear stochastic equation cannot be solved exactly. We are thus forced to restrict the discussion to moderate shear rates. 
But this does \emph{not} mean that $\dot{\gamma} \tau_c$ should remain small, as might be expected from simple scaling arguments.
Indeed, our study reveals that the parameter that actually governs the steady state properties of the system is the combination of the two
dimensionless parameters
\begin{equation}
\alpha = \sqrt{ \frac{k_BT}{\sigma l_c^2}}  \times \dot{\gamma}_{\mathit{eff}} \tau_c\ .
\label{alpha}
\end{equation}
Depending on experimental conditions, $\alpha$ can be small even if $\dot{\gamma} \tau_c$ is not~\cite{derksPRL06}.
The analysis presented in the following is valid in the range $0 \leq \alpha < 1$.

\begin{figure}
\includegraphics[width=80mm]{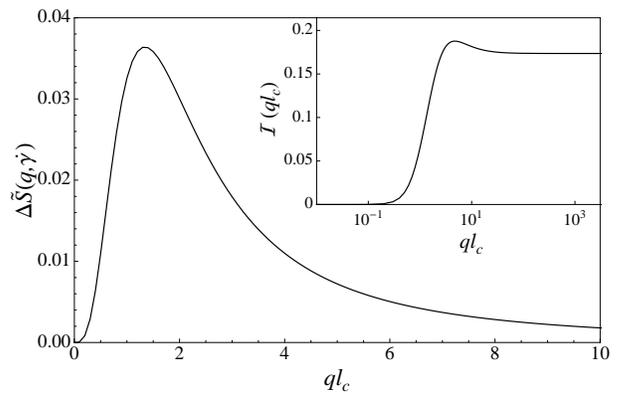}
\caption{Correction to the nonequilibrium fluctuation spectrum  $\Delta \tilde{S}(\mathbf{q},\dot{\gamma})
= \sigma /(k_B T l_c^2) \times \Delta S(\mathbf{q},\dot{\gamma})$,  in the direction of shear $\theta_q =0$ and for $\alpha =1$.  In inset we show $\mathcal{I}(ql_c)$. }
\label{sdeq}
\end{figure}

Eq.~(\ref{mceq}) is solved using a perturbation theory presented in Appendix~\ref{app4}.
We first discuss the steady-state correlation function $S(\mathbf{q},\dot{\gamma})$
  defined as 
 \begin{equation}
\lim_{t \rightarrow \infty} \langle h (\mathbf{q},t) h(\mathbf{q}',t)\rangle   =
 S(\mathbf{q},\dot{\gamma}) (2\pi)^2 \delta ( \mathbf{q}+\mathbf{q}')   \ .
 \label{power}
 \end{equation}
It is given at equilibrium by $S(\mathbf{q},0)=k_BT /(\sigma( q^2 + l_c^{-2}))$. 
Under shear, the spectrum is modified according to
 \begin{equation}
 S(\mathbf{q},\dot{\gamma}) =  S(\mathbf{q},0) \left[ 1 -\alpha^2 \mathcal{I}(ql_c) \cos^2 \theta_q  + \mathcal{O} (\alpha^4) \right] \ ,
 \label{sqshear}
 \end{equation}
with $\theta_q$ the angle between the direction of shear $\mathbf{e}_x$ and the wave vector $\mathbf{q}$.
The function $\mathcal{I}(ql_c)$ depends only on the norm $q=\vert \mathbf{q} \vert$ of the wave vector. Explicitly,  we get
 \begin{eqnarray}
\mathcal{I} (x) =  \displaystyle{\frac{1}{\pi^2}}  \int \d^2 \mathbf{s} \cos \theta_s     
  \frac{x^2 s   }{\vert \mathbf{x}-\mathbf{s} \vert f ( \vert \mathbf{x}-\mathbf{s} \vert )}  \nonumber \\
 \times \displaystyle{ \frac{  \left[xf(x)\right]^{-1} -\left[sf(s)\right]^{-1} }{f(x)+f(s)+f ( \vert \mathbf{x}-\mathbf{s} \vert )}} \ ,
\label{integral}
 \end{eqnarray}
with $x=\vert \mathbf{x} \vert$, $s=\vert \mathbf{s} \vert$, $\theta_s$ the angle between $\mathbf{x}$ and $\mathbf{s}$, and $f(x)=(1+x^2)/x$. 
This integral  cannot be evaluated analytically; the result of the numerical integration is presented in the inset of Fig.~\ref{sdeq}~\cite{remnum}.

Eq.~(\ref{sqshear}) is the second main result of this paper.
It shows that the coupling is maximum in the flow direction ($\theta_q=0$), while the spectrum is not affected in the vorticity direction ($\theta_q=\frac{\pi}{2}$).
As expected, the result is invariant by inversion symmetry $q_x \leftrightarrow -q_x$ or  $q_y \leftrightarrow -q_y$.
The correction $\Delta S(\mathbf{q},\dot{\gamma})=\left\vert S(\mathbf{q},\dot{\gamma})-S(\mathbf{q},0) \right\vert$
is plotted in Fig.~\ref{sdeq} in dimensionless units. It should be noticed that even though all the wave lengths are affected by the flow, 
the spectrum is mostly affected when $q \sim l_c^{-1}$. The correction then vanishes  like $q^{-2}$ for larger values of $q$,
whereas  it scales as $\Delta S(\mathbf{q},\dot{\gamma})\sim q^4$ in the limit $q \rightarrow 0$.

The mean square displacement of the interface is then obtained from the sum over all  modes
 $\langle h^2 \rangle (\dot{\gamma}) = (2\pi )^{-2} \int S(\mathbf{q},\dot{\gamma}) q 
\d q  \d \theta_q  $. Notice that the integral that defines the roughness at equilibrium is actually  divergent. Regularization is achieved by means of a microscopic cut-off $a$ so that $\langle h^2 \rangle (0)=k_BT/(2\pi \sigma)\times \ln \left( l_c/a \right)$. In NESS, we find that the fluctuations are strongly reduced by the external flow
\begin{equation} \label{hcarre}
\langle h^2 \rangle (\dot{\gamma}) = \langle h^2 \rangle (0) \left[ 1 -K \alpha^2 +  \mathcal{O} (\alpha^4) \right]  \ .
 \end{equation}
The correction is quadratic in the control parameter $\alpha$.
In this expression, $K$ is a \emph{universal} constant in the sense that it depends neither on the properties of the fluids nor of the elastic constants of the 
interface~\cite{rem2}. Moreover, it is independent of the microscopic cut-off as soon as $a/l_c <10^{-2}$. 
Numerically, we get $K \approx 0.087$.

\begin{figure}
\includegraphics[width=80mm]{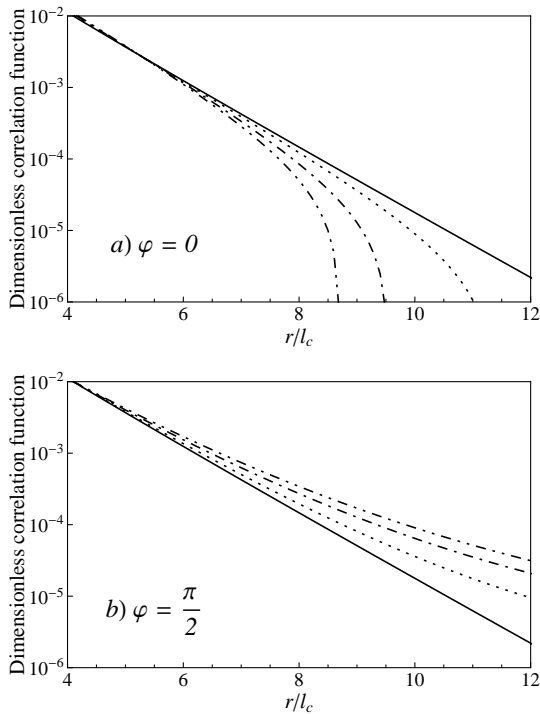}
\caption{Dimensionless correlation function $2\pi \sigma/(k_BT ) \times C(\mathbf{r},\dot{\gamma}) $ for different values of the control parameter: $\alpha=0$ (solid line), $\alpha=0.5$ (dot), $\alpha=0.8$ (dash-dot), and $\alpha=1$ (dash-dot-dot). a) In the direction of shear ($\varphi=0$), the correlation length decreases. b) In the vorticity direction ($\varphi=\pi /2$), the correlation length increases. }
\label{correlfig}
\end{figure}

Finally, we focus our attention on the NESS correlation function $C(\mathbf{r},\dot{\gamma})= \langle h(\mathbf{r},t)h(\mathbf{0},t) \rangle$, which is the Fourier transform of the structure factor. As can be seen in Eq.~(\ref{sqshear}), the spectrum $S(\mathbf{q},\dot{\gamma})$ is not modified for the modes that are perpendicular to the flow
(\emph{i.e.} wave vectors with  $q_x = 0$). However, performing the sum over all wave vectors gives rise, in direct space, to modifications
\emph{even in the vorticity direction}.
Denoting $\varphi$ the angle between the direction of shear $\mathbf{e}_x$ and the position vector $\mathbf{r}$, the correlation function reads
 \begin{eqnarray}
C(\mathbf{r},\dot{\gamma})= \frac{k_BT}{2\pi \sigma} \big[ K_0 \left( r/l_c \right) - \alpha^2 \cos^2  \varphi \, C_x\left( r/l_c \right) \nonumber \\
- \alpha^2 \sin^2  \varphi \, C_y\left( r/l_c \right)  \big] \ ,
\label{correl}
 \end{eqnarray}
with $K_0$ the modified Bessel function of the second kind. The corrections can be written $C_x(x)=C_1(x)-C_2(x)/x$ and $C_y(x)=C_2(x)$,
where for $i=1,2$
 \begin{equation}
C_i(x) =  \int_0^{\infty} \d s  \frac{J_{i-1} (sx)}{1+s^2} \mathcal{I}(s)  s^{2-i}  \ .
 \end{equation}
Here, $J_n$ is the Bessel function of the first kind.

At equilibrium, $C(\mathbf{r},0)$ is proportional to  $K_0 \left( r/l_c \right)$ and behaves like  $C(\mathbf{r},0) \propto \exp[-r/l_c]$ at large separations $r \gg l_c$. In other words the capillary length is also the equilibrium correlation length.  In nonequilibrium situation, the correlation function  presents a number of interesting features. We first consider the direction of shear  $\varphi=0$. It clearly appears from Fig.~\ref{correlfig}.a) that $C(\mathbf{r},\dot{\gamma})$ decays faster that its equilibrium counterpart. This reflects a \emph{decrease} of the correlation length as the control parameter $\alpha$ is increased.  The situation is different in the vorticity direction $\varphi=\pi/2$, as shown on Fig.~\ref{correlfig}.b). In this case the decay of $C(\mathbf{r},\dot{\gamma})$ is slower than the equilibrium correlation function, indicating an \emph{increase} of the correlation length with increasing shear rate.

Since we do not have an explicit functional form of the correlation function, it is difficult  to be more quantitative regarding the correlation length.  The only conclusion that can be drawn from Fig.~\ref{correlfig} is that the decay of $C(\mathbf{r},\dot{\gamma})$ at large distances is not exponential anymore.
Finally note that at intermediate angles ($0< \varphi <\pi/2$),  the correlation length  can be either larger or smaller than $l_c$ depending on the shear rate.

\section{Discussion}
\label{disc}

Complex fluids under shear is an challenging class of nonequilibrium systems. The simple system considered in this work is a fluid interface exposed to a Couette flow.
First, it was shown that the time evolution of the deformation satisfies a mode-coupling equation.
Note that equations similar to Eq.~(\ref{mceq}) have been suggested in the context of soft surfaces  under shear  (see for instance~\cite{marlowPRE02,brayPRE01a,brayPRE01b}). 
As a matter of fact, it belongs to the general class of the Kardar-Parisi-Zhang equation~\cite{kardarPRL86} whose derivation is usually based on phenomenological grounds. Here however,  Eq.~(\ref{mceq}) has been \emph{rigorously} derived from hydrodynamics with no other assumption that inertial effects can be neglected.
We now discuss our results in view of recent studies.

\subsection{Comparison to experiments and simulations}

\begin{figure}
\includegraphics[width=80mm]{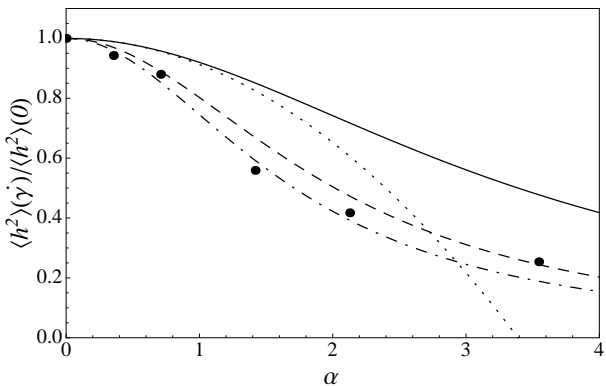}
\caption{Mean square displacement of the interface a a function of the control parameter $\alpha$. The solid circles refer to the experiment of Ref.~\cite{derksPRL06} (sample A). The dotted and solid lines are the theoretical predictions of Eqs.~(\ref{hcarre}) and~(\ref{hcarre2}), respectively, with $K_{\mathit{theo}} \approx 0.087$.
The dashed line corresponds to the fit of Eq.~(\ref{hcarre2}) with $K_{\mathit{fit}}\approx 0.246$.
The dash-dotted line is the model of Derks \emph{et al.}~\cite{derksPRL06}.}
\label{msd}
\end{figure}

We have shown  that interfacial fluctuations are strongly reduced by the flow.
According to Eq.~(\ref{hcarre}), the reduction is governed by a universal parameter: 
we predict $K \approx 0.087$. Suppression of thermal capillary waves was indeed measured by Derks and collaborators in a recent experiment~\cite{derksPRL06,derksphd}.  
This was achieved by using a phase-separated colloid-polymer mixture whose interface is characterized by a very low surface tension.
The authors have used two compositions but the dynamical parameters are only available for the first sample (sample A, very close to  the critical point). 
The thickness of the fluid layers are  $L_1=50$~$\mu$m  and $L_2=350$~$\mu$m. Equilibrium parameters obtained from the correlation functions are $\sigma = 2.5$~nN/m, $l_c=2.6$~$\mu$m, and $\tau_c=13$~s~\cite{derksPRL06}. The viscosities extracted from the velocity profiles are $\eta_1= 7.8$~mPa.s and $\eta_2=5.2$~mPa.s~\cite{derksphd}.
Unfortunately, there are only two experimental points (out of 6) that correspond to the condition $\alpha <1$.
To fit the data for  higher values of the control parameter,
we assume that Eq.~(\ref{hcarre}) can be extended  according to
\begin{equation} 
\langle h^2 \rangle (\dot{\gamma})  \simeq  \langle h^2 \rangle (0) \frac{1}{  1 + K \alpha^2   }  \ .
\label{hcarre2}
 \end{equation}
We then perform a linear regression of the quantity $\langle h^2 \rangle (0)/\langle h^2 \rangle (\dot{\gamma})$ as a function of $\alpha^2$.
We find $K_{\mathit{fit}}\approx 0.246$, almost 3 times the expected value. The comparison between our theoretical predictions and the experimental measurements 
is  shown on  Fig.~\ref{msd}. As can be seen the agreement is only qualitative, even though it is difficult to be really conclusive since most data are outside the validity range of  Eq.~(\ref{hcarre})~\cite{rem3}.

In Ref.~\cite{derksPRL06}, the suppression of thermal capillary waves  is supported by a remarkably simple model.
The authors state that only the slow modes with $\dot{\gamma} \tau_q >1$ are affected by the flow.
Solving this inequality defines to two wave vectors $q_1$ and $q_2$. It is then assumed that wave vectors with $q_1<q<q_2$ no longer contribute to the fluctuation spectrum, whereas the contribution from wave vectors outside this range is left unchanged. Their prediction of the interfacial roughness as a function of the applied shear rate $\dot{\gamma}$  is in very good agreement with experimental data -- see Fig.~\ref{msd}. Still, we argue that the hydrodynamic analysis presented in this paper provides new insights into the problem. 
In particular, we have shown that  fluctuations are driven out of equilibrium by the effective shear rate $\dot{\gamma}_{\mathit{eff}}$ rather than the applied shear rate  $\dot{\gamma}$. For sample~A presented above, the ratio of effective to applied shear rate is  $\dot{\gamma}_{\mathit{eff}}/\dot{\gamma} \approx 0.8 $.
Now, suppose that the experiment is done with the same sample composition excepted that  the thicknesses of the fluid layers 
are inverted to $L_1=350$~$\mu$m  and $L_2=50$~$\mu$m. 
In this case, the ratio would be $\dot{\gamma}_{\mathit{eff}}/\dot{\gamma} \approx 1.2 $: for the same value of the applied shear rate, $\alpha$ is thus expected to increase by 50~$\%$.
The ensuing difference regarding the mean square displacement should be substantial enough for the distinction between applied and effective shear rate to be experimentally observed.
This would provide a direct validation of our analysis.

Finally, let us discuss the correlation functions. Our study reveals that the static structure factor is strongly modified in the flow direction, whereas fluctuation modes in the vorticity direction are not affected. This result confirms similar conclusions obtained from molecular dynamic simulations~\cite{thakreJCP08}. Coming back in real space, we find that the correlation length decreases in the direction of shear. This conclusion agrees with Monte Carlo simulations of a driven interface~\cite{smithPRL08}. Physically, it would mean that the external flow acts as an effective potential whose strength should increase with the shear rate. However, these findings are in disagreement with the results of Derks \emph{et al.} that observe an increase of the correlation length~\cite{derksPRL06}.
Note that experimental data are fitted using the equilibrium correlation function. 
This procedure is mainly based on the short distance behavior of $C(r,\dot{\gamma})$.
In contrast, our conclusions are drawn  from the \emph{asymptotic} behaviour of the correlation function. 
It is thus possible that the discrepancy comes from how the correlation length is defined in practice.
Other possibilities are discussed below.

\subsection{Conjectures and future directions}

The equation of motion~(\ref{mceq}) is a direct generalization of the dispersion relation for capillary waves. In particular, it is based on the same assumptions:  continuum hydrodynamics and small-gradient expansion.
The pertinent question is  whether the apparent mismatch between theoretical predictions and  experimental data can be the consequence of another mechanism~\cite{rem4}.
In fact, several  explanations  are plausible.

Firstly, we remark that the diameter of the colloids is of the order of $140$~nm. The capillary length is thus only ten or twenty times larger than the ``microscopic'' cut-off, and it is not clear whether the separation of length scales assumed to get the universal quantities ($K$, $\mathcal{I}$, \ldots)  is really achieved for this sample.

Secondly, Monte Carlo simulations of Smith \emph{et al.}  reveals that the average \emph{width} of the interface is significantly reduced by the flow~\cite{smithPRL08}. 
In the capillary wave theory, it is assumed that the density profile behaves like a step function with a strictly vanishing width.
But it is known that physical parameters  are directly related to the shape of the \emph{true} density profile~\cite{rowlinsonbook}.
For instance, the surface tension is given by
\begin{equation} 
\sigma = m \int \d z  \left( \frac{ \partial \rho}{\partial z} \right)^2  \ ,
\label{defsigma}
 \end{equation}
with $\rho(z,0)$ the equilibrium profile and $m$ a constant proportional to the second moment of the intermolecular potential~\cite{rowlinsonbook}.
In the small but finite interfacial region, the density is varying rapidly so that the fluids are highly compressible. The density should therefore be modified by the flow and become a function of the applied shear rate $\rho(z,\dot{\gamma})$. 
A plausible consequence is that the bare surface tension might  depend on $\dot{\gamma}$ as well. 
According to Eq.~(\ref{defsigma}), the thinning of the average profile observed in~\cite{smithPRL08}  is expected to give rise to an increase of the surface tension,
and then to an additional reduction of the fluctuations.

The latter assertion is certainly hypothetical but
we argue that the measurement of the correlation function in the \emph{vorticity} direction could  be conclusive.
Indeed, an increase of the bare surface tension  would  lead to an increase of $l_c$  and thus of the NESS correlation length. 
But a modification of $C(\mathbf{r},\dot{\gamma})$ arising from  $\sigma (\dot{\gamma})$ is expected to be \emph{isotropic},
whereas the hydrodynamic theory predicts a modification  that is \emph{anisotropic}.
This question clearly deserves to be reconsidered both from the experimental and theoretical viewpoint.

\section{Conclusion}
\label{ccl}

Let us briefly summarize the main results of this paper.

(i) We have generalized the hydrodynamic derivation of the dispersion relation  to include the effect of an applied shear rate on the dynamics of capillary waves. 
Eq.~(\ref{mceq}) reveals that the system is driven out of equilibrium by an effective shear rate that differs from the applied one.
The mode-coupling structure of the equation also shows that nonequilibrium fluctuations cannot be described in terms of a simple renormalization of the interfacial tension.

(ii) The analysis of the mode-coupling equation shows that the fluctuations are smoothed out by the flow, in qualitative agreement with experiments
and simulations. We find that NESS fluctuations can be described in terms of a single control parameter defined in Eq.~(\ref{alpha}). Conclusions regarding the height correlation function are more questionable. If the theory and the simulations predict a decrease of the correlation length in the direction of shear, the reverse is actually observed on the  experimental side.  The discrepancies might be due to the finite thickness of the interfacial region and its response to the external flow. 
Molecular dynamics simulations of interfaces under shear should help to clarify this point in a near future.

In conclusion, we have derived a hydrodynamic theory to account for the coupling between a shear flow and the fluctuations of an interface.
Previous  investigations on driven interfaces have revealed partial agreements but also significant discrepancies between experiments and numerical simulations.
Although our work brings new insights into the issue, it seems that further investigations will be necessary to definitely reconcile experiment, theory and simulations. Ultimately, achievement of this program should allow to reach a deeper understanding of soft matter systems under shear.

\acknowledgments

The authors wish to thank D.~Aarts, D.~Derks, E.~Rapha\"el, M.~Schmidt and A.~W\"urger for stimulating discussions. We are also indebted to D.~Derks for providing us with experimental data. 
The Institute of Fundamental Physics (IPF) -- Bordeaux is acknowledged for financial support in the early stages of this work.

\appendix

\section{Derivation of the boundary conditions}
\label{app1}

Since the analysis presented in this work involves nonlinear contributions, particular care has to be given to the conditions at the interface. 
In this appendix, we first remind the derivation of the kinematic relation Eq.~(\ref{kinematic}). We then focus on the stress continuity condition Eq.~(\ref{stress}).

\subsection{Kinematic condition}

It is convenient to represent the interface with the functional $f(x,y,z,t)=z-h(x,y,t)$. 
The unit vector $\mathbf{n}$ normal to the surface is given by 
\begin{equation}
\mathbf{n} = \frac{ \bm{\nabla} f}{ \vert  \bm{\nabla} f \vert}  \ .
\label{normalapp}
\end{equation}
Let us focus on a point $M(x,y,z)$ that belongs to the interface. It satisfies $f(x,y,z,t)=0$ and then
\begin{equation*}
\d f = \partial_t f \d t + \bm{\nabla} f \cdot \d \mathbf{r} = 0  \ .
\end{equation*}
If $\mathbf{w}$ is the velocity of the point $M$, the infinitesimal displacement 
reads $\d \mathbf{r} = \mathbf{w} \d t$ and therefore $\bm{\nabla} f \cdot  \mathbf{w} =  -  \partial_t f $. Together with Eq.~(\ref{normalapp}), we then obtain
\begin{equation*}
\mathbf{w} \cdot  \mathbf{n} = - \frac{\partial_t f}{\vert  \bm{\nabla} f \vert} \ .
\end{equation*}
Identifying the normal velocity of the interface $\mathbf{w} \cdot  \mathbf{n}$ with the normal velocity of the fluid $\mathbf{v} \cdot  \mathbf{n}$, one finally gets
 \begin{equation}
 \frac{ \partial_t h}{\sqrt{1+ (\bm{\nabla} h)^2}}  = \mathbf{v} \cdot  \mathbf{n}  \ .
 \end{equation}
Here, we have also used the fact that $\partial_t f = - \partial_t h$.
This relation together with the definition~(\ref{normal}) directly leads to Eq.~(\ref{kinematic}).

\subsection{Dynamical condition}

To derive the stress condition~(\ref{stress}), we
consider an infinitesimal surface element $\mathcal{S}$ of the interface, bound by a closed contour $\mathcal{C}$. The contour is chosen circular with radius $\epsilon$. We also define a small cylinder $\mathcal{V}$, centered on the contour $\mathcal{C}$ and with height $2 \epsilon$. The forces acting on the volume element $\mathcal{V}$ are the body force density $\mathbf{f}$, the tension force exerted along the perimeter $\mathcal{C}$, and the surface force exerted by the fluids above and below the cylinder. We discuss separately each contribution.

\subsubsection{Tension force}

We first discuss the restoring force exerted on the contour line $\mathcal{C}$. We assume that the contour is travelled counterclockwise. The (local) orthonormal basis associated with the line is denoted $\{ \mathbf{m},\mathbf{s},\mathbf{n} \}$, with $\mathbf{m}$ the tangent vector, $\mathbf{s}$ the vector normal to $\mathcal{C}$ but tangent to $\mathcal{S}$, and $\mathbf{n}$ the vector normal to $\mathcal{S}$ pointing towards the upper fluid. In this representation, the external force exerted on the contour $\mathcal{C}$ reads
\begin{equation*}
\mathbf{F}_t = - \oint \sigma \mathbf{s} \d l = \sigma \oint \left( \mathbf{m} \times \mathbf{n} \right) \d l 
= \sigma \oint \left( \d \mathbf{l} \times \mathbf{n} \right)   \ ,
\end{equation*} 
with $\d \mathbf{l} =  \mathbf{m} \d l$. Then, according to the Stokes relation  $\oint  \d \mathbf{l} \times \mathbf{A}= \int \left( \d \mathbf{S} \times \bm{\nabla} \right) \times \mathbf{A}$, one gets
\begin{equation*}
\mathbf{F}_t = \sigma  \int \d S \left( \mathbf{n} \times \bm{\nabla} \right) \times \mathbf{n} \ ,
\end{equation*}
with the surface element $\d \mathbf{S} = \d S \mathbf{n}$. 

Finally, using the general identity
$\left(\mathbf{A} \times \bm{\nabla}\right) \times \mathbf{A} =\frac{1}{2} \bm{\nabla} \left(\mathbf{A} \cdot \mathbf{A}\right) - \mathbf{A} \left(\bm{\nabla} \cdot \mathbf{A}\right)$ and because $ \mathbf{n} \cdot \mathbf{n} =1$, we find 
\begin{equation}
\label{ft}
\mathbf{F}_t = - \sigma  \int \d S \left(\bm{\nabla} \cdot \mathbf{n}\right) \,  \mathbf{n}  \ .
\end{equation}

\subsubsection{Surface force}

The component $T_{\mu \nu}$ of the stress tensor $\mathsf{T}$ corresponds the $\mu^{th}$ component of the force (per unit area) acting on a surface element normal to the direction $\nu$. In particular, the force density exerted by a fluid on a surface element with outward normal $\hat{\mathbf{n}}$ is $ \mathsf{T} \cdot \hat{\mathbf{n}}$.

Let us apply this definition to the elementary cylinder of volume $\mathcal{V}$. The cylinder is sitting across the surface, the extension being $\epsilon$ on each side. Regarding the upper surface of the cylinder, the stress exerted by fluid 1 is characterized by the (outside) normal vector $\hat{\mathbf{n}}_1=  \mathbf{n}$ so that  the force density reads $ \mathsf{T}_1 \cdot \mathbf{n}$.
Similarly on the lower surface one has $\hat{\mathbf{n}}_2= - \mathbf{n}$ and the stress  is $-\mathsf{T}_2 \cdot \mathbf{n}$. The total surface force is then
\begin{equation}
\mathbf{F}_s = \int \d S \left( \mathsf{T}_1 - \mathsf{T}_2  \right) \cdot \mathbf{n} \ .
\end{equation}

\subsubsection{Force balance equation}

The force balance on the volume element $\mathcal{V}$ enclosing the surface $\mathcal{S}$ defined by the contour $\mathcal{C}$
then reads
\begin{equation*}
\int_\mathcal{V} \d V \rho \frac{ \d \mathbf{v} }{\d t }  =  \mathbf{F}_s + \mathbf{F}_t   +  \int_\mathcal{V} \d V  \mathbf{f} \ .
\end{equation*}
Now if $\epsilon$ is the typical lengthscale of the volume element $\mathcal{V}$, then the surface contributions scale as $\epsilon^2$ whereas the acceleration and the body forces scale as $\epsilon^3$. Hence in the limit 
$\epsilon \rightarrow 0$ the latter can be neglected and  surface forces must balance. However, since the surface element is arbitrary, the integrand must vanish identically so that one finally gets the desired result
\begin{equation}
\label{bcstress}
\left( \mathsf{T}_1 - \mathsf{T}_2  \right) \cdot \mathbf{n} = \sigma \left(\bm{\nabla} \cdot \mathbf{n} \right) \,  \mathbf{n}  \ . 
\end{equation}
Finally notice that the divergence of the normal vector reads in the Monge representation~\cite{kamienRMP02}  
\begin{equation}
\label{curv}
\bm{\nabla} \cdot \mathbf{n}  = - \nabla^2 h + \mathcal{O}(h^3)  \ . 
\end{equation}

\section{Nonlinear relaxation equation}
\label{app2}

The aim of this section is to derive the equation of motion~(\ref{mceq}).
The fluctuation modes of the interface are expected to be deformed by the shear flow. As we shall see, the description of this effect requires to go 
beyond the usual linear analysis. We focus here on the relaxation dynamics in the absence of noise. Thermal fluctuations are treated in Appendix~\ref{app3}.

To begin with, we recall the Stokes equation
 \begin{equation}
 \eta_i \nabla^2 \mathbf{v} - \bm{\nabla} p + \rho_i \mathbf{g}   = \mathbf{0}  \ ,
 \label{stokesapp}
 \end{equation}
with $\mathbf{g} = -g \mathbf{e}_z$ the gravitational acceleration. In the small-gradient approximation, we can express the deformation profile as $h(x,y,t) = \varepsilon u(x,y,t)$ with  $u(x,y,t) \sim \mathcal{O}(1)$. 
The dimensionless parameter that governs the small-gradient expansion is 
 \begin{equation}
\varepsilon = \sqrt{\frac{k_BT }{\sigma l_c^2}} \ .
\label{epsapp}
\end{equation}
We then assume that the fields that satisfy the Stokes equation can be expanded in powers of $\varepsilon$
 \begin{equation}
F(z) = F^{(0)}(z)  + \varepsilon F^{(1)}(z) + \varepsilon^2 F^{(2)}(z) + \mathcal{O} \left( \varepsilon^3 \right)  \ ,
\label{devapp}
 \end{equation}
where $F$ stands for $\mathbf{v}$, $\mathsf{t}$ or $p$. 
(For clarity reason, only the $z$-dependence of the fields is  kept explicitly in this appendix).
The solution at order $\varepsilon^0$ satisfies Eq.~(\ref{stokesapp}); it corresponds to the Couette flow solution for a flat interface.  
The other terms of the expansion~(\ref{devapp}) are solution of the following equation
 \begin{equation}
 \eta_i \nabla^2 \mathbf{v}^{(n)} - \bm{\nabla} p^{(n)}    = \mathbf{0}  \ , \ \forall n\geq 1 \ .
 \label{stokesnapp}
 \end{equation}
At order $\varepsilon^1$, the solution in the absence of shear would lead to the usual dispersion relation for capillary waves. Here, the calculations are carried out up to order $\varepsilon^2$.  

Due to the linearity of Eq.~(\ref{stokesnapp}), the coupling between different orders only occurs through the boundary conditions at the interface. This suggests to use a recursive method in order to solve the problem~\cite{bickelEPJE02}.

\subsection{Solution at order $\varepsilon^0$: flat interface}

For a flat interface, the solution is the well-known solution for the Couette flow:
 \begin{eqnarray}
& \mathbf{v}^{(0)} (z)  =& \dot{\gamma}_i z \mathbf{e}_x  \label{v0}  \\
&p^{(0)} (z) =& P^{(0)} - \rho_i g z  \label{p0}
 \end{eqnarray}
with $i=1$ if $z \geq 0$ and $i=2$ if $z\leq 0$. $P^{(0)}$ is a constant. The stress tensor then reads
 \begin{equation}
 \mathsf{t}^{(0)}(z) =
\begin{pmatrix}
- p^{(0)}(z) & 0 & \eta_i \dot{\gamma}_i \\
0 & - p^{(0)}(z) & 0 \\
\eta_i \dot{\gamma}_i  & 0  & - p^{(0)}(z)
\end{pmatrix}
\label{t0}
 \end{equation}
In particular, the continuity relation~(\ref{stress}) for the tangential stress implies $ \eta_1 \dot{\gamma}_1 = \eta_2 \dot{\gamma}_2 $.

Note that we have chosen $L_1$, $V_1$, $L_2$ and $V_2$ so that the plane of zero shear is the horizontal plane $z=0$.
At constant shear rate, any other situation can be deduced thanks to a simple Galilean transformation.

\subsection{Solution at order $\varepsilon^1$: linear relaxation}

The first-order solution satisfies Eq.~(\ref{stokesnapp}) with boundary conditions that  involve the solution at order $\varepsilon^0$. 
Actually, there are two kinds of conditions that needs to be enforced: continuity conditions at the interface and limit conditions far from the interface.
Strictly speaking, the latter have to be expressed at the boundaries of the shear cell, $z \rightarrow L_1$ or $z \rightarrow -L_2$, where $L_1$ and $L_2$ are \emph{macroscopic} lengths. In this work, we are interested on thermal fluctuations that develop on \emph{microscopic} lengthscales (i.e.,  with wavelength much smaller than $L_1$ or $L_2$). We can therefore assume that for any field $F$,  the terms of the series~(\ref{devapp}) with $n \geq 1$ vanish at infinity
 \begin{equation}
\lim_{z \rightarrow \pm \infty} F^{(n)}(z)=0   \ , \ \forall n\geq 1 \ .
\label{limit}
 \end{equation}
In the following, we first derive the boundary conditions at the interface and then solve the problem in Fourier representation.

\subsubsection{Boundary conditions}

To express the continuity conditions, let us define for a field $F$ the notation 
 \begin{equation}
\left[ F \right]_{z_0} = F(z_0^+)-F(z_0^-)
 \end{equation}
with $F(z_0^{\pm}) = \lim_{z \rightarrow z_0} F(z)$, the limit being taken for $z\gtrless z_0$.
The first condition that needs to be enforced is the continuity of the velocity at the interface 
 \begin{equation*}
\left[  \mathbf{v} \right]_h= \mathbf{0}  \ .
 \end{equation*}
Using the expansion~(\ref{devapp}) together with the Taylor expansion of $\mathbf{v}(h=\varepsilon u)$, we obtain at first order
 \begin{equation*}
\left[ \mathbf{v}^{(1)}  + u \partial_z  \mathbf{v}^{(0)} \right]_0  = \mathbf{0}  \ .
 \end{equation*}
From the result~(\ref{v0}) we get $\partial_z \mathbf{v}^{(0)}=\dot{\gamma}_i \mathbf{e}_x$, and finally
\begin{subequations}
 \begin{eqnarray}
\left[  v_x^{(1)} \right]_0 &=& \left(\dot{\gamma}_2-\dot{\gamma}_1  \right) u  \ ,  \label{velxapp} \\
\left[  v_y^{(1)} \right]_0 &=& 0  \ ,  \label{velyapp} \\
\left[  v_z^{(1)} \right]_0 &=& 0  \ . \label{velzapp}
 \end{eqnarray}
 \label{velocityapp}
 \end{subequations}

Next, we have to enforce the condition of stress continuity Eq.~(\ref{stress}). At first order, we obtain 
\begin{equation*}
\left[t_{ \mu z }^{(1)} + u \partial_z t_{\mu z}^{(0)}\right]_0  
 = - \sigma \nabla^2 u \delta_{ \mu z }
\end{equation*}
with $\mu=x,y$ or $z$. Following the result~(\ref{t0}) for the stress tensor, we find
\begin{equation*}
\partial_z t_{ \mu z}^{(0)} (z)= \rho_i g \delta_{ \mu z}  \ .
\end{equation*}
The continuity conditions for the  stress tensor then reads
\begin{subequations}
\begin{eqnarray}
\left[t_{xz}^{(1)}\right]_0 &= & 0 \ , \label{stressxapp}  \\
\left[t_{yz}^{(1)}\right]_0 &= & 0 \ ,  \label{stressyapp}   \\
\left[t_{z z}^{(1)} \right]_0  &=& \Delta \rho g u - \sigma \nabla^2 u  \ ,  \label{stresszapp} 
\end{eqnarray}
\label{stressapp}
\end{subequations}
with  $\Delta \rho = \rho_2-\rho_1$.

\subsubsection{Method and solution}

The solution of the problem is expressed using  the 
two-dimensional Fourier transform 
 \begin{equation}
 \begin{cases}
F (\mathbf{q},z,t)=\displaystyle{\int \d^2 \mathbf{r}\exp [-i\mathbf{q} \cdot \mathbf{r}]F(\mathbf{r},z,t)}  \ , \\
F (\mathbf{r},z,t)=\displaystyle{\int \frac{\d^2 \mathbf{q}}{(2\pi)^2} \exp [-i\mathbf{q} \cdot \mathbf{r}]F(\mathbf{q},z,t)}  \ ,
\end{cases}
 \label{fourier}
 \end{equation}
with $\mathbf{r}=(x,y)$ and $\mathbf{q}=(q_{x},q_{y})$.
The difficulty of the algebra comes from the fact that we need to solve the full
3-dimensional problem.
It then appears judicious  to
define a new orthogonal coordinate system that would account for the 
symmetries of the problem. To this aim,
the vector fields are decomposed into their
longitudinal, transverse and normal components~\cite{bickelEPJE06,bickelPRE07}.
This defines a new set of orthogonal unit vectors
$(\mathbf{l},\mathbf{t},\mathbf{e}_z)$
where $\mathbf{l}$ is the unit vector parallel to $\mathbf{q}$,
and $\mathbf{t}$ the in-plane vector perpendicular to
$\mathbf{l}$ and $\mathbf{e}_z$.
These vectors are expressed in the cartesian basis
$(\mathbf{e}_x,\mathbf{e}_y,\mathbf{e}_z)$ as
\begin{align}
\mathbf{l} &=\frac{q_x}{q}\mathbf{e}_x +  \frac{q_y}{q}\mathbf{e}_y  \ , \nonumber \\
\mathbf{t} &= - \frac{q_y}{q}\mathbf{e}_x +  \frac{q_x}{q}\mathbf{e}_y    \ . \nonumber
\end{align}
The velocity is then written as 
$ \mathbf{v}^{(1)} = v_l^{(1)} \mathbf{l}
 +v_t^{(1)} \mathbf{t}
+v_z^{(1)}  \mathbf{e}_z$.
Inserting this expression in Eq.~(\ref{stokesnapp})
leads to a  system of differential equations
for the Fourier-transformed quantities
\begin{subequations}
\begin{align}
-\eta_{i} q^2 v_l^{(1)}  + \eta_{i} \partial_z^2 v_l^{(1)}   &=  i q p^{(1)}   \ ,  \label{stokesl} \\
-\eta_{i} q^2 v_t^{(1)} + \eta_{i} \partial_z^2 v_t^{(1)}   &= 0   \label{stokest}  \ ,    \\
-\eta_{i} q^2 v_z^{(1)} + \eta_{i} \partial_z^2 v_z^{(1)}  &=  \partial_z p^{(1)}  \ ,  \label{stokesn}
\end{align}
\label{stokesappfourier}
\end{subequations}
together with the divergenceless condition
\begin{equation}
i q v_l^{(1)} + \partial_z v_z^{(1)} =0 \ .
\label{incompressibility}
\end{equation}

It is not difficult to show that the solution of Eqs.~(\ref{stokesappfourier}) and~(\ref{incompressibility}) can be written
\begin{subequations}
\label{solo1}
\begin{align}
v_z^{(1)}(z) &= A_i^{(1)} e^{-q \vert z \vert } + B_i^{(1)} q  z  e^{-q \vert z \vert }   \ , \\
v_t^{(1)}(z) & = C_i^{(1)} e^{-q \vert z \vert }   \ , \\
p^{(1)}(z) &= 2 \eta_i  B_i^{(1)}  q e^{-q \vert z \vert }  \ .
\end{align}
\end{subequations}
$A_i^{(1)}$, $B_i^{(1)}$, and $C_i^{(1)}$ are the (yet undetermined) integration constants, the subscript $i$ taking the value $i=1$ if $z \geq 0$ and $i=2$ if $z \leq 0$. (There should be \emph{a priori}
no confusion between the subscript  and the imaginary unit~$i$). The solution for the longitudinal component follows directly from~(\ref{incompressibility}).
Note that  the condition~(\ref{limit}) has already been taken into account in~(\ref{solo1}).

Next, we have to express the boundary conditions Eqs.~(\ref{velocityapp}) 
and~(\ref{stressapp}) in Fourier space and in the new system of coordinates. The condition~(\ref{velzapp}) for the normal component is straightforward
\begin{equation}
\left[  v_z^{(1)} \right]_0 = 0  \ ,
\label{vzq}
\end{equation}
where of course $ v_z^{(1)}(\mathbf{q},z,t)$  now stands for  the Fourier transformed quantity. Then projecting conditions~(\ref{velxapp}) and~(\ref{velyapp}) for the parallel components onto the directions $\mathbf{l}$ and $\mathbf{t}$ respectively gives
\begin{equation}
\left[  v_l^{(1)} \right]_0 = \frac{q_x}{q}  \left(\dot{\gamma}_2-\dot{\gamma}_1  \right) u  \ ,
\label{vlq}
\end{equation}
and 
\begin{equation}
\left[  v_t^{(1)} \right]_0 = \frac{q_y}{q}  \left(\dot{\gamma}_1-\dot{\gamma}_2  \right) u  \ .
\label{vtq}
\end{equation}
Regarding the continuity of the stress, the condition~(\ref{stresszapp}) is now expressed as
\begin{equation}
\left[  - p^{(1)} + 2 \eta_i \partial_z v_z^{(1)} \right]_0 = \sigma \left(q^2+ l_c^{-2} \right) u  \ ,
\label{szq}
\end{equation}
where we define the capillary length by $l_c^2 = \sigma / (\Delta \rho g)$.
Finally, the projection of~(\ref{stressxapp}) and~(\ref{stressyapp}) leads to
\begin{equation}
\left[  \eta_i \left(  \partial_z v_l^{(1)} +iq v_z^{(1)} \right) \right]_0 = 0  \ ,
\label{slq}
\end{equation}
and
\begin{equation}
\left[  \eta_i \partial_z v_t^{(1)}  \right]_0 =0  \ .
\label{stq}
\end{equation}

With the boundary conditions~(\ref{vzq}) -- (\ref{stq}), we can now evaluate the 6 integration constants.
We find
\begin{subequations}
\label{inconst1}
\begin{align}
 A_1^{(1)} &  = A_2^{(1)} =- \frac{1}{4 \bar{\eta} q}  \sigma \left(q^2+ l_c^{-2} \right) u  \ , \label{a11} \\
B_1^{(1)}  &=  A_1^{(1)} +  i\frac{q_x}{2 \bar{\eta} q}  \dot{\gamma}_1 \left(\eta_2-\eta_1 \right)u   \ , \\
B_2^{(1)}  &=  -A_2^{(1)} +  i\frac{q_x}{2 \bar{\eta} q}  \dot{\gamma}_2 \left(\eta_1-\eta_2 \right)u  \ , \\
C_1^{(1)}  &=  \frac{q_y}{2 \bar{\eta} q}  \dot{\gamma}_1 \left(\eta_2-\eta_1 \right)u  \ , \\
C_2^{(1)}  &=  \frac{q_y}{2 \bar{\eta} q}  \dot{\gamma}_2 \left(\eta_1-\eta_2 \right)u  \ , 
\end{align}
\end{subequations}
with $\bar{\eta} = (\eta_1 + \eta_2)/2$. 

Although the first-order solution already depends on the shear flow, it is claimed in the main body of the paper that the calculations
have to be performed up to second order. To understand this point, consider  the kinematic condition~(\ref{kinematic}).
Since $\mathbf{v}^{(0)}(0)=\mathbf{0}$, we can write at  first order
\begin{equation*}
\partial_t h = \varepsilon v_z^{(1)}(0) = -  \frac{1}{\tau_q} h  \ ,
\end{equation*}
with $\tau_q = -1/A_1^{(1)}$.
It appears from~(\ref{a11}) that the shear rate is not involved in the equation of motion at linear order. The calculations have thus to be extended to include the first nonlinear contribution.

\subsection{Solution at order $\varepsilon^2$: mode-coupling equation}

\subsubsection{General solution}

The solution at second order satisfies the same set of linear differential equations~(\ref{stokesappfourier}) and~(\ref{incompressibility}), so that it
takes the form
\begin{subequations}
\label{solo2}
\begin{align}
v_z^{(2)}(z) &= A_i^{(2)} e^{-q \vert z \vert } + B_i^{(2)} q  z  e^{-q \vert z \vert }   \ , \\
v_t^{(2)}(z) & = C_i^{(2)} e^{-q \vert z \vert }   \ , \\
p^{(2)}(z) &= 2 \eta_i  B_i^{(2)}  q e^{-q \vert z \vert }  \ .
\end{align}
\end{subequations}
Again, $v_l^{(2)}$ is deduced from the incompressibility condition~(\ref{incompressibility}).

It appears that not all the integration constants are needed for our purpose. Indeed, let us express the kinematic condition~(\ref{kinematic})
up to second order. In direct space, it reads
\begin{align*}
\varepsilon \partial_t u + \varepsilon^2  u \partial_x u & \partial_z v_x^{(0)}(0) + \varepsilon^2 \bm{\nabla}_{\parallel} u \cdot \mathbf{v}_{\parallel}^{(1)}(0)  \\
& = \varepsilon    v_z ^{(1)}(0) + \varepsilon^2 u  \partial_z  v_z^{(1)}(0) + \varepsilon^2 v_z ^{(2)}(0) \ ,
\end{align*}
where we have used the fact that $\mathbf{v}^{(0)}(0) = \mathbf{0}$. Note that the passage from~(\ref{kinematic}) to this latter equation leads to an
indeterminacy. Indeed, although the velocity field $\mathbf{v}$ is continuous at $z=h$, this is not the case for each term of the expansion~(\ref{devapp}) at $z=0$.
In this appendix, we thus assume that the limit $z \rightarrow 0$ is taken from above.  It can be checked that the same results are obtained if one would take the limit from below. We now switch to Fourier space, where the product of two functions $f(\mathbf{r}). g(\mathbf{r})$ is transformed into the convolution
product
\begin{equation}
\left( f \ast g \right) (\mathbf{q}) \doteq \int \frac{\d^2 \mathbf{k}}{(2\pi)^2} f \left( \mathbf{k} \right) g \left(  \mathbf{q} -\mathbf{k} \right) \ .
\end{equation}
We then have in Fourier representation
\begin{align*}
\varepsilon \partial_t u  &= \varepsilon    v_z ^{(1)}(0) + \varepsilon^2 \Big[  \big( u \ast  \partial_z  v_z^{(1)}(0) \big) - i \dot{\gamma}_1 \big(  u \ast q_x u \big)  \\
&  - i  \big( v_x^{(1)}(0) \ast q_x u \big)  
 - i   \big( v_y^{(1)}(0) \ast q_y u \big)  \Big]  +  \varepsilon^2v_z ^{(2)}(0) \ .
\end{align*}
With the results of the preceding section, the equation of motion can now be written as
\begin{equation}
\varepsilon \partial_t u  = - \frac{1}{\tau_q}\varepsilon   u - i \varepsilon^2 \dot{\gamma}_1 \big(u \ast q_x u \big) \left[ \frac{\eta_1 - \eta_2}{\bar{\eta}} +1 \right]  
+  \varepsilon^2   A_1^{(2)}   \ .
\label{motionapp}
\end{equation}
Therefore, only the integration constant $A_1^{(2)}$ is required to get the equation of motion.

\subsubsection{Integration constants}

At second order, the continuity condition~(\ref{velocity}) for the $z$-component of the velocity is expressed in direct space as
 \begin{equation*}
\left[  v_z^{(2)}  + u \partial_z  v_z^{(1)} + \frac{u^2}{2}\partial^2_z  v_z^{(0)} \right]_0  = 0 \ ,
 \end{equation*}
leading to,  in reciprocal space,
 \begin{equation}
\left[  v_z^{(2)}   \right]_0  = i  \left( \dot{\gamma}_2 - \dot{\gamma}_1 \right)  \big(u \ast q_x u \big) \ .
\label{vz2}
 \end{equation}
This relation actually involves both $A_1^{(2)}$ and $A_2^{(2)}$. To obtain one last equation, we still need to express the normal stress condition
 \begin{equation*}
\big[  \mathbf{n} \cdot \mathsf{t} \cdot \mathbf{n}   \big]_h  = - \sigma \nabla^2 h + \mathcal{O} \left( h^3 \right) \ .
 \end{equation*}
Together with~(\ref{t0}) and~(\ref{stressapp}),  it reads at  order $\varepsilon^2$
 \begin{equation*}
\left[   t^{(2)}_{zz} + u \partial_z   t^{(1)}_{zz} \right]_0  = 0 \ .
 \end{equation*}
This result is finally expressed in Fourier representation. The derivative of the stress tensor at first order is then obtained from Eq.~(\ref{stokesn})
 \begin{align*}
 \partial_z   t^{(1)}_{zz}  & = - \partial_z p^{(1)} + 2 \eta_i \partial_z^2  v^{(1)}_{z}  \\
 & = \eta_i \left(  q^2 v^{(1)}_{z} +  \partial_z^2  v^{(1)}_{z}  \right)  \\
 & = \eta_i \left( q^2  v^{(1)}_{z} -iq \partial_z  v^{(1)}_{l}  \right) \ ,
 \end{align*}
where we have used~(\ref{incompressibility}) to get the last equality. From the continuity relation~(\ref{slq}), it is not difficult 
to obtain the condition
 \begin{equation}
\left[   t^{(2)}_{zz}\right]_0  = 0 \ ,
 \end{equation}
that leads to the simple relation
 \begin{equation}
\eta_1 A_1^{(2)} + \eta_2 A_2^{(2)}  = 0 \ .
 \end{equation}
Inserting this result in Eq.~(\ref{vz2}), we finally get
\begin{subequations}
\begin{align}
 A_1^{(2)} = i  \dot{\gamma}_1 \big(u \ast q_x u \big) \left( \frac{\eta_1 - \eta_2}{2\bar{\eta}} \right) \ , \\
 A_2^{(2)} = i  \dot{\gamma}_2 \big(u \ast q_x u \big) \left( \frac{\eta_2 - \eta_1}{2\bar{\eta}} \right) \ .
\end{align}
\end{subequations}
If we factorize all  second-order contributions in~(\ref{motionapp}), we then arrive at the equation of motion 
\begin{equation}
 \partial_t \varepsilon u  = - \frac{1}{\tau_q}\varepsilon   u - i   \frac{2 \eta_1 \dot{\gamma}_1}{\eta_1 + \eta_2} \big(\varepsilon u \ast q_x \varepsilon u \big)   \ .
\label{motionapp1}
\end{equation}
Coming back to the deformation field $h = \varepsilon u$ and using the fact that $\eta_1 \dot{\gamma}_1 = \eta_2 \dot{\gamma}_2$, we can  write the
equation of motion in its definitive form Eq.~(\ref{mceq}).

\section{Fluctuating hydrodynamics}
\label{app3}

In this appendix, we consider the Landau-Lifshitz equation of linear fluctuating hydrodynamics~\cite{ortizbook}. 
We thus need to adapt the perturbative scheme developed in Appendix~\ref{app2} to the Stokes equation~(\ref{stokes})
including the random part of the stress tensor $\mathsf{s}$. To this aim, it should be noticed that the noise correlations
$ \langle s_{\mu \nu} s_{\mu' \nu'} \rangle$
given in Eq.~(\ref{randomten}) scale as $ k_BT $
and are therefore proportional to  $\varepsilon^2$ [see Eq.~(\ref{epsapp})].
From this viewpoint,  the random part of the stress tensor  can be identified as a   \emph{first-order contribution}.

We then follow the same recursive method as in Appendix~\ref{app2}. There is no modification at order $\varepsilon^0$ and we now discuss the first and second orders of the expansion~(\ref{devapp}).

\subsection{Solution at order $\varepsilon^1$}

At first order, the stochastic version of 
the Stokes equation~(\ref{stokes}) reads, in the $(\mathbf{l},\mathbf{t},\mathbf{e}_z)$ basis,
\begin{subequations}
\begin{align}
\eta_{i} \left( \partial_z^2 -   q^2 \right) v_l^{(1)}    &=  i q p^{(1)} -iq s_{ll} - \partial_z s_{lz}   \ ,  \\
\eta_{i} \left( \partial_z^2 -   q^2 \right) v_t^{(1)}   &= - \partial_z s_{tz}     \ ,    \\
\eta_{i} \left( \partial_z^2 -   q^2 \right) v_z^{(1)}  &=  \partial_z p^{(1)} -iq s_{lz} - \partial_z s_{zz}   \ ,  
\end{align}
\label{fluctstokes}
\end{subequations}
together with the incompressibility condition
\begin{equation}
i q v_l^{(1)} + \partial_z v_z^{(1)} =0 \ .
\label{fluctinc}
\end{equation}
As explained in the text, the fluctuations of the random forces in the bulk are not affected by the shear flow (at least in the regime considered in this paper).
The correlations are then given by the fluctuation-dissipation theorem~\cite{landaups2,ortizbook}. In this representation, we have
 \begin{align}
\left\langle s_{\mu \nu} s_{\mu' \nu'}   \right\rangle  
  &=  2  k_B T \eta_i   \left( \delta_{\mu \mu'}\delta_{\nu \nu'} + \delta_{\mu \nu'}\delta_{\nu \mu'} \right)    \nonumber \\
&\times  (2\pi)^2 \delta \left( \mathbf{q} + \mathbf{q}'  \right) \delta (z-z') \delta (t-t')  \ ,
\label{fluctcor}
 \end{align}
where the subscripts stand for $l$, $t$ or $z$.

The solution of Eqs.~(\ref{fluctstokes}) and~(\ref{fluctinc}) is then obtained  using the following Green's function identity~\cite{grantPRA83}
 \begin{equation*}
\Big[ \partial^2_z -q^2 \Big] \frac{1}{2q} e^{-q\vert z-z' \vert } = - \delta (z-z')  \ .
 \end{equation*}
Regarding the boundary conditions, the stress tensor $\mathsf{t}$ in~(\ref{velocityapp}) and~(\ref{stressapp}) has now to be replaced by the \emph{total}
stress tensor $\mathsf{T} = \mathsf{t} + \mathsf{s}$~\cite{haugeJSP73,grantPRA83}. After some algebra, we obtain the evolution equation for the interface
 \begin{equation}
 \partial_t h (\mathbf{q},t)=- \frac{1}{\tau_q} h(\mathbf{q},t)  + \varphi (\mathbf{q},t)   \ ,
 \end{equation}
where the noise  $\varphi (\mathbf{q},t)$ is given by
 \begin{eqnarray}
 \varphi (\mathbf{q},t) = \frac{q}{4 \bar{\eta} } \int_0^{+ \infty}  \d z z e^{-qz} \left( s_{zz} - s_{ll} +2i s_{lz} \right)  \nonumber \\
+ \frac{q}{4 \bar{\eta} } \int_{- \infty}^0  \d z z e^{qz} \left( s_{zz} - s_{ll} -2i s_{lz} \right) \ .
\label{appcorphi}
 \end{eqnarray}
Together with the correlations~(\ref{fluctcor}), we finally get the result announced in  Eq.~(\ref{noisecorrel}). Note that Eq.~(\ref{appcorphi}) generalizes the result of 
Grant and Desai for a liquid-gas interface (in the absence of shear)~\cite{grantPRA83}.

\subsection{Solution at order $\varepsilon^2$}

At second order, the velocity field satisfies the following set of equations
\begin{subequations}
\begin{align}
\eta_{i} \left( \partial_z^2 -   q^2 \right) v_l^{(2)}    &=  i q p^{(2)}   \ ,  \\
\eta_{i} \left( \partial_z^2 -   q^2 \right) v_t^{(2)}   &=0     \ ,    \\
\eta_{i} \left( \partial_z^2 -   q^2 \right) v_z^{(2)}  &=  \partial_z p^{(2)}  \ ,  
\end{align}
\end{subequations}
together with the continuity equation
\begin{equation}
i q v_l^{(2)} + \partial_z v_z^{(2)} =0 \ .
\end{equation}
The noise  enters the problem only through the boundary conditions and the first order contribution.
Again, the boundary conditions are the same as in Appendix~\ref{app3} excepted that the stress tensor $\mathsf{t}$
 is replaced by the total stress tensor $\mathsf{T}=\mathsf{t}+ \mathsf{s}$.

The calculations are pretty lengthy so that we only give the final results. The equation of motion is obtained from the kinematic condition Eq.~(\ref{kinematic}).
Up to second order, we find
\begin{widetext}
 \begin{equation}
 \partial_t h   = -  \frac{1}{\tau_q}  h  - i   \dot{\gamma}_{\mathit{eff}} \big[ h \ast  \left(q_x h \right) \big] 
 +   \varphi  - \mathbf{q} \cdot \big[  \left(\mathbf{q} \phi_1\right)    \ast h \big] - \Big(  \mathbf{q}  \times \big[  \left(\mathbf{q} \phi_2\right)    \ast h \big] \Big) \cdot \mathbf{e}_z  \ ,
 \end{equation}
 \end{widetext}
where $\phi_1$ and $\phi_2$ are defined as
 \begin{align}
 \phi_1 (\mathbf{q},t) &= \frac{1}{4 \bar{\eta}q } \int_0^{+ \infty}  \d z \left( 1 -qz \right) e^{-qz} \left( s_{zz} - s_{ll} +2i s_{lz} \right)  \nonumber \\
&+ \frac{1}{4 \bar{\eta}q } \int_{- \infty}^0  \d z \left(1+qz \right) e^{qz} \left( s_{zz} - s_{ll} -2i s_{lz} \right) \ ,
 \end{align}
and 
 \begin{eqnarray}
 \phi_2 (\mathbf{q},t) = \frac{1}{2 \bar{\eta} q} \int_0^{+ \infty}  \d z e^{-qz} \left(  s_{lt} +i s_{tz}  \right)  \nonumber \\
+ \frac{1}{2 \bar{\eta}q } \int_{- \infty}^0  \d z  e^{qz} \left(  s_{lt} -i s_{tz}  \right) \ .
 \end{eqnarray}

It can be noticed that the additional terms involving $\phi_1$ and $\phi_2$ are not induced by the external flow. At equilibrium, these second-order contributions  
would actually renormalize the relaxation dynamics of the interface. In nonequilibrium situation, it is argued in the next appendix
that those terms do not contribute to the fluctuation spectrum at the order considered in this work. Therefore, we do not to mention them in the main document. This choice is made in order to improve the readability and to focus the discussion on the main outcomes.

\section{Solution of the mode-coupling equation}
\label{app4}

\subsection{Linear noise}

Eq.~(\ref{mceq}) is solved using a perturbation theory. To this aim, let us define the time Fourier transform $\tilde{f} (\mathbf{q},\omega ) $ of any function $f(\mathbf{q},t)$ as $\tilde{f} (\mathbf{q},\omega )=\int \d t \exp [-i \omega t ]f(\mathbf{q},t) $. The mode-coupling equation can then be rewritten 
 \begin{align}
 \tilde{h} (\mathbf{q},\omega)    = &    \tilde{R}_0 (\mathbf{q},\omega)  \tilde{\varphi}   (\mathbf{q},\omega)   - i\dot{\gamma}_{\mathit{eff}}
  \tilde{R}_0 (\mathbf{q},\omega)  \nonumber \\
 & \times  \int \d \underline{\mathbf{k}}     k_x \tilde{h}(\mathbf{k}  ,\Omega) \tilde{h}(\mathbf{q-k},\omega - \Omega)  \ ,
 \end{align}
with $\d \underline{\mathbf{k}}  =\d^2 \mathbf{k}  \d \Omega /(2 \pi)^3$. The bare propagator $\tilde{R} (\mathbf{q},\omega)$ is given by
 \begin{equation}
\tilde{R} (\mathbf{q},\omega) = \frac{\tau_q}{1 + i \omega \tau_q}  \ .
 \end{equation}
The solution of the problem is  then obtained as the solution of a Dyson equation. The calculations are performed up to second order in 
the small parameter $\alpha$ (which is proportional to $ \dot{\gamma}_{ \mathit{eff} } $). Using the shortland notation 
$\tilde{h}_{\mathbf{q}} = \tilde{h} (\mathbf{q},\omega)  $, we get
\begin{widetext}
 \begin{align}
 \tilde{h}_{\mathbf{q}}   =     \tilde{R}_{\mathbf{q}}  \tilde{\varphi}_{\mathbf{q}}  - i\dot{\gamma}_{\mathit{eff}}
  \tilde{R}_{\mathbf{q}}    \int \d \underline{\mathbf{k}}  &  k_x  \tilde{R}_{\mathbf{k}}  \tilde{R}_{\mathbf{q}-\mathbf{k}} \tilde{\varphi}_{\mathbf{k}}  \tilde{\varphi}_{\mathbf{q}-\mathbf{k}}  
  - \dot{\gamma}_{\mathit{eff}}^2 \tilde{R}_{\mathbf{q}}   \int \d \underline{\mathbf{k}} \d \underline{\mathbf{k}}'    k_x k'_x  \tilde{R}_{\mathbf{k}}   \tilde{R}_{\mathbf{q}-\mathbf{k}} \tilde{R}_{\mathbf{k}'}   \tilde{R}_{\mathbf{q}-\mathbf{k}'}   \tilde{\varphi}_{\mathbf{q}-\mathbf{k}} \tilde{\varphi}_{\mathbf{k}'} \tilde{\varphi}_{\mathbf{q}-\mathbf{k}'} \nonumber    \\
 & - \dot{\gamma}_{\mathit{eff}}^2 \tilde{R}_{\mathbf{q}}   \int \d \underline{\mathbf{k}} \d \underline{\mathbf{k}}'    k_x k'_x  \tilde{R}_{\mathbf{k}}   \tilde{R}_{\mathbf{q}-\mathbf{k}} \tilde{R}_{\mathbf{k}'}  \tilde{R}_{\mathbf{q}-\mathbf{k}-\mathbf{k}'}   \tilde{\varphi}_{\mathbf{k}} \tilde{\varphi}_{\mathbf{k}'} \tilde{\varphi}_{\mathbf{q}-\mathbf{k}-\mathbf{k}'}  + \mathcal{O} \left(\alpha^3 \right)  \ ,
 \label{appdyson}
 \end{align}
 \end{widetext}
From this expansion, the evaluation of the correlation function $\langle \tilde{h}_{\mathbf{q}}  \tilde{h}_{\mathbf{q}'}    \rangle$
is pretty lengthy but presents no conceptual difficulty since the four-point correlation functions of the noise are given by the Wick's theorem. The result is then Fourier-transformed back in time representation, leading to Eqs.~(\ref{sqshear}) and~(\ref{integral}).

\subsection{Nonlinear noise}

The same scheme is used to treat the nonlinear noise contributions. First, we need to specify the statistical properties of $\phi_1$ and $\phi_2$.
From~(\ref{fluctcor}), it is not difficult to get $\langle \phi_1 \rangle = \langle \phi_2 \rangle =0$, and for $\mu$, $\nu=1$ or $2$,
 \begin{equation}
\langle \phi_{\mu} (\mathbf{q},t)\phi_{\nu} (\mathbf{q}',t')\rangle =
 \frac{k_BT}{2\bar{\eta} q^3 }   \delta (t-t')  (2\pi)^2 \delta ( \mathbf{q}+\mathbf{q}') \delta_{\mu \nu}   \ ,
 \end{equation}
and
 \begin{equation}
\langle \phi_{\mu} (\mathbf{q},t)\varphi (\mathbf{q}',t')\rangle = 0  \ .
 \end{equation}
The solution is then expressed as a Dyson equation similar to Eq.~(\ref{appdyson}). Although the expression of $ \tilde{h}_{\mathbf{q}} $ involves more terms, it can be shown that most of the additional contributions to  $\langle  \tilde{h}_{\mathbf{q}}   \tilde{h}_{\mathbf{q}'}  \rangle$ actually vanish (either because they involve an odd number of noise terms, or because of the vanishing cross-correlations between $\varphi$, $\phi_1$ and $\phi_2$).
As a matter of fact, additional shear-induced contributions are at least of order $\varepsilon^6$, and thus do not contribute to the modification of the spectrum at the order considered in this work.

Still, the nonlinear noise \emph{does} give rise to non-vanishing contributions at order $\varepsilon^4$. But those terms are \emph{equilibrium} contributions since they do not involve the shear rate. Consider for instance the mean square displacement: those terms would simply renormalize the equilibrium  value $\langle h^2 \rangle (0)$. This is why we have carefully plotted the \emph{ratio} $\langle h^2 \rangle (\dot{\gamma}) / \langle h^2 \rangle (0)$, that represents the modification of the fluctuations by the shear with respect to the \emph{equilibrium} situation~\cite{rem5}.


\begin{thebibliography}{0}
\bibitem{rem1}
As a matter of fact, our analysis reveals that convolutions between deformation and noise of the form $\int \d \mathbf{q}' h(\mathbf{q}',t) s_{\mu \nu} ( \mathbf{q}-\mathbf{q}',t) \ldots$ also occur in the equation of motion (see Appendix~\ref{app3}). These nonlinear terms  are not coupled to the flow. At equilibrium, they would actually renormalize the relaxation dynamics of the interface  However, it is explained in Appendix~\ref{app4} that they do not contribute to the fluctuation spectrum at order $\varepsilon^4$, so that they will not be considered in the following.
\bibitem{rem2}
Note that from this viewpoint, the integral $\mathcal{I}$ is also a universal quantity.
\bibitem{rem3}
Actually, the solution of Eq.~(\ref{mceq}) can be extended to high values of the shear rate following for instance the self-consistent approach detailed
in~\cite{schwartzJSM08}. However, the results are very similar to those of  the approxiation~(\ref{hcarre2}) so that we do not present more details.
\bibitem{rem4}
Note that the theory admits no free parameter.
\bibitem{rem5}
As a matter of fact, the modification of the equilibrium relaxation dynamics would deserve a study on its own.
\end{thebibliography}
\end{document}